\documentclass[prl,twocolumn,showpacs]{revtex4}

\usepackage{graphicx}
\usepackage{amsmath}

\begin{document}

\title{$p$-Wave Interactions in Low-Dimensional Fermionic gases}

\author{Kenneth G{\"u}nter, Thilo St{\"o}ferle, Henning Moritz, Michael K{\"o}hl$^\dag$,  and
Tilman Esslinger}

\affiliation{Institute of Quantum Electronics, ETH Z\"{u}rich,
H\"{o}nggerberg, CH--8093 Z\"{u}rich, Switzerland}

\date{\today}

\begin{abstract}

We study a spin-polarized degenerate Fermi gas interacting via a
$p$-wave Feshbach resonance in an optical lattice. The strong
confinement available in this system allows us to realize one- and
two-dimensional gases and therefore to restrict the asymptotic
scattering states of atomic collisions. When aligning the atomic
spins along (or perpendicular to) the axis of motion in a
one-dimensional gas, scattering into channels with the projection of
the angular momentum of $|m|=1$ (or $m=0$) can be inhibited. In two
and three dimensions we observe the doublet structure of the
$p$-wave Feshbach resonance. Both for the one-dimensional and the
two-dimensional gas we find a shift of the position of the resonance
with increasing confinement due to the change in collisional energy.
In a three-dimensional optical lattice the losses on the Feshbach
resonance are completely suppressed.

\end{abstract}


\maketitle

Ultracold fermionic atoms constitute a well controllable
many-particle quantum system which provides access to fundamental
concepts in physics. Using optical lattices the atomic motion and
the dimensionality of the trapping geometry can be controlled. Yet,
it is the collisional interaction between atoms which provides the
avenue towards the physical richness of the strongly correlated
regime \cite{Hofstetter2002,Rigol2003,Jaksch2005,Koehl2005}.
Particularly intriguing are $p$-wave collisions due to their
anisotropic character. These can be experimentally accessed
exploiting a Feshbach resonance which overcomes the suppression of
the collisional cross section at ultralow energies
\cite{Regal2003,Zhang2004,Schunck2005}.

In this paper we prepare a spin-polarized Fermi gas in an optical
lattice and investigate $p$-wave collisions in the vicinity of a
Feshbach resonance controlled by the magnetic field. We study the
resonant behavior of the atom losses as a function of the magnetic
field and observe distinct features depending on the
dimensionality and the symmetry of the system. For a
three-dimensional gas a double-peaked structure appears, as has
previously been reported by Ticknor {\it et al.}\
\cite{Ticknor2004}. This characteristic survives when the
dimensionality is reduced to two dimensions (2D) but appears
shifted in magnetic field. For one-dimensional (1D) geometries
only a single shifted resonance peak is observed. All resonantly
enhanced losses vanish when the spin-polarized gas is loaded into
the lowest band of a three-dimensional optical lattice, in which
each site can be regarded as a system of ''zero dimensions``.

\begin{figure}[htbp]
  \includegraphics[width=.5\columnwidth,clip=true]{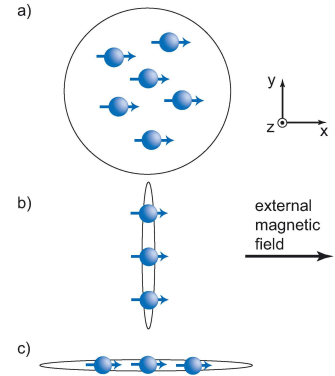}
  \caption{Spin-alignment dependent interactions in 1D and 2D. In the
  two-dimensional configuration of a) all projections of the
  angular momentum in the $p$-wave collision are allowed.
  b) and c) show a one-dimensional spin-polarized Fermi gas
  with the spins aligned orthogonal and parallel to the
  extension of the gas, respectively. In b) only the $|m|=1$ projection of the $p$-wave
  contributes to the scattering, in c) only the $m=0$ projection.}
  \label{fig1}
\end{figure}

\begin{figure}[htbp]
  \includegraphics[width=.75\columnwidth,clip=true]{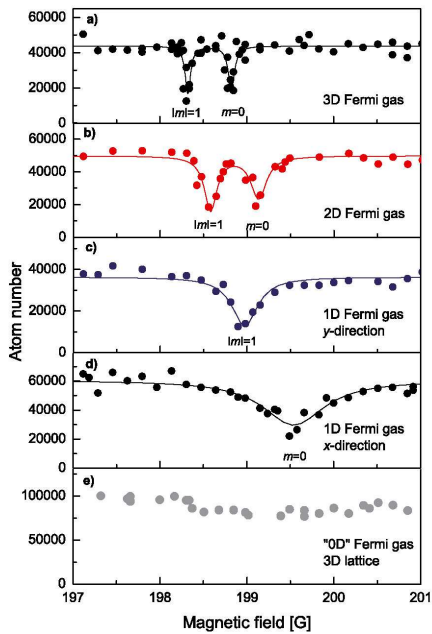}
  \caption{Loss measurements of the $p$-wave Feshbach resonance. a) Atoms are held in a
  crossed-beam optical dipole trap. b) Two-dimensional Fermi gas ($V_z=25\,E_r$).
  c) One-dimensional Fermi gas with the motion confined orthogonal to
  the direction of the magnetic field ($V_z=V_x=25\,E_r$). d) One-dimensional Fermi gas with the motion confined parallel
  to the direction of the magnetic field ($V_z=V_y=25\,E_r$). e) Fermi gas in a three-dimensional optical lattice ($V_x=V_y=V_z=25\,E_r$)
  The solid lines are Lorentzian fits to the data from which we extract
  the position and the width of the resonance.}
  \label{fig2}
\end{figure}

These observations can be qualitatively explained by considering
the symmetry of the collisions, as illustrated in Fig.\
\ref{fig1}. The external magnetic field orients the polarization
of the atoms and its direction may be chosen as the quantization
axis. In order to describe the atom-atom scattering with $p$-wave
symmetry, the angular part of the corresponding asymptotic wave
functions can be expressed in terms of spherical harmonics.
Alignment of the scattering state parallel to the quantization
axis corresponds to the spherical harmonic $Y_{\ell=1,m=0}$ and
alignment in the plane perpendicular to the quantization axis
corresponds to superpositions of the spherical harmonics
$Y_{\ell=1,m=\pm 1}$. The dipole-dipole interaction between the
electronic spins lifts the degeneracy between the $|m|=1$ and the
$m=0$ collisional channels which leads to a splitting of the
Feshbach resonance \cite{Ticknor2004}. In the two- and
three-dimensional configurations both collisional channels are
present, giving rise to the observed doublet feature (see Fig.\
\ref{fig2}a and b). In one dimension, with the spin aligned
orthogonal or parallel to the atomic motion, either the $|m|=1$
(see Fig.\ \ref{fig2}c) or the $m=0$ (see Fig.\ \ref{fig2}d)
collisional channel is contributing, leading to a single peak. In
''zero dimensions'' -- as realized in a three-dimensional optical
lattice -- $p$-wave collisions and the corresponding losses are
absent (see Fig.\ \ref{fig2}e). In these low-dimensional systems
the asymptotic scattering states are kinematically restricted.
However the atomic collision process is still three-dimensional
since the size of the ground state is large compared to the range
of the interatomic potentials. Hence the strongly confined
directions can contribute to the collision energy \cite{Kim2001}.

% History
One- and two-dimensional fermionic quantum systems have been
realized in semiconductor nanostructures \cite{Fowler1966} and
recently with noninteracting \cite{Modugno2003} and interacting
\cite{Jochim2003,Moritz2005} atomic gases in optical lattices. In
these systems the strong confinement modifies the scattering
properties of the particles: It stabilizes molecular states and
shifts the position of Feshbach resonances. This has been
predicted for one- \cite{Olshanii1998,Bergeman2003} and
two-dimensional systems \cite{Petrov2000,Wouters2003} interacting
via $s$-wave scattering, and confinement induced molecules have
been observed in a 1D gas \cite{Moritz2005}. Similarly, for
spin-polarized Fermions in one dimension a confinement induced
shift of $p$-wave Feshbach resonances is predicted
\cite{Granger2004}.

% Experimental procedure
Our experimental procedure used to produce a degenerate Fermi gas
has been described in detail in previous work \cite{Koehl2005}.
Fermionic $^{40}$K atoms are sympathetically cooled by thermal
contact with bosonic $^{87}$Rb atoms, the latter being subjected
to forced microwave evaporation. After reaching quantum degeneracy
for both species with typically $6\times 10^5$ potassium atoms in
the $|F=9/2, m_F=9/2\rangle$ hyperfine state at a temperature of
$T/T_F=0.35$ ($T_F$ is the Fermi temperature), we remove all
rubidium atoms from the trap. The potassium atoms are then
transferred from the magnetic trap into an optical dipole trap
consisting of two intersecting laser beams along the horizontal
$x$- and $y$-directions. These laser beams are derived from diode
lasers at a wavelength of $\lambda=826\,$nm and are focused to
$1/e^2$-radii of $50\,\mu$m ($x$-axis) and $70\,\mu$m ($y$-axis).
In the optical trap we prepare the atoms in the $|F=9/2, m_F=-7/2
\rangle$ spin state at a magnetic bias field of 232.9\,G using two
radio frequency (rf) sweeps. To remove residual atoms in the
$|F=9/2, m_F=-9/2 \rangle$ state we change the magnetic field in
100\,ms to a value of 201.7\,G, close to the $s$-wave Feshbach
resonance between $|F=9/2,m_F=-9/2\rangle$ and
$|F=9/2,m_F=-7/2\rangle$ \cite{Regal2003,Moritz2005}, where we
encounter inelastic losses resulting in a pure spin-polarized
Fermi gas. Subsequently we increase the magnetic field within
100\,ms to 203.7\,G. Then we evaporate atoms by lowering the
optical trapping potential during 2.5\,s to a final value of
$7\,E_r$ in each of the two beams, where $E_r=\hbar^2
k^2/(2m_\text{K})$ denotes the recoil energy, $k=2\pi/\lambda$ the
wave vector of the laser and $m_\text{K}$ the atomic mass. The
preparation of the gas is completed by rapidly ($< 1$\,ms)
decreasing the magnetic field to 194.4\,G, which is below the
$p$-wave Feshbach resonance. We have calibrated the magnetic field
by rf spectroscopy between Zeeman levels with an accuracy better
than 100\,mG, and we estimate the reproducibility of our magnetic
fields to be better than 50\,mG.

% Reference: Crossed-Beam dipole trap
For comparison with the low-dimensional situations we first study
the $p$-wave Fesh\-bach resonance in the crossed-beam optical trap
where motion in all three dimensions is possible. We sweep the
magnetic field from its initial value of 194.4\,G using a linear
ramp within 1\,ms to its final value in the vicinity of the
Feshbach resonance. There the atoms are subject to inelastic
losses \cite{Regal2003}. After a hold time of 6.4\,ms we switch
off both the magnetic field and the optical trap and let the
atomic cloud expand ballistically for 7\,ms before we take an
absorption image. From the image we extract the remaining number
of atoms. In these data (see Fig.\ \ref{fig2}a) we observe the
doublet structure of the $p$-wave Feshbach resonance, which is due
magnetic dipole-dipole interactions between the atoms
\cite{Ticknor2004}. The decay constant of the atom number close to
the Feshbach resonance is on the order of 1\,ms, which is
comparable to the settling time of the magnetic field. Therefore
we encounter a systematic shift on the order of +0.1\,G due to the
direction of the magnetic field ramp \cite{shift}.

% 2D Fermi system
In a next step, we additionally apply a single optical standing wave
along the vertical $z$-axis. The standing wave with a potential
depth $V_z$ \cite{calibrationlattice} creates a stack of
two-dimensional Fermi gases in the horizontal $x$-$y$-plane. The
lattice laser intensity is increased using an exponential ramp with
a time constant of 10\,ms and a duration of 20\,ms. The beam for the
vertical optical lattice is derived from a diode laser at a
wavelength of $\lambda=826$\,nm and is focused to a $1/e^2$-radius
of $70\,\mu$m. The magnetic field is aligned along the horizontal
$x$-axis, as depicted in Fig.\ \ref{fig1}a. In the two-dimensional
Fermi gas we have studied the $p$-wave Feshbach resonance analogous
to the method described above, only the release process of the atoms
is slightly altered: within 1\,ms before the simultaneous switch-off
of the magnetic and the optical potentials, we lower the lattice
intensity to $V_z=5\,E_r$ to reduce the kinetic energy. This results
in a more isotropic expansion which allows to determine the atom
number more precisely.

% Shift in 2D Part 1
For the two-dimensional gas we observe a similar doublet structure
of the Feshbach resonance but shifted towards higher magnetic
field values with respect to the position without strong
confinement (see Fig.\ \ref{fig2}b). Due to the angular momentum
in a $p$-wave collision there is a centrifugal barrier in addition
to the interatomic potential, which results in a pronounced energy
dependence of the scattering. In the confined gas the collision
energy is modified by the motional ground state energy and the
larger Fermi energy of the gas due to the confinement. Moreover, a
confinement induced shift of the resonance could be envisaged,
similar to what has been studied for $s$-wave interactions in two
dimensions \cite{Petrov2000,Wouters2003}.

\begin{figure}[htbp]
  \includegraphics[width=.77\columnwidth,clip=true]{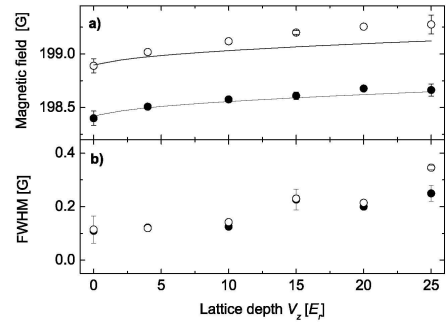}
  \caption{a) Shift of the Feshbach resonance position when
  tuning the gas from three to two dimensions. Open symbols indicate the position of the $m=0$ branch,
  solid symbols the $|m|=1$ branch of the resonance. The error bars denote the statistical error of 3 measurements. The solid
  lines show a calculation of the expected positions (see text). b) Evolution of the full width at half maximum (FWHM) of the loss feature.}
  \label{fig3}
\end{figure}

% Shift in 2D Part 2
We experimentally find that the shift of the resonance feature
depends on the strength of the optical lattice. In Fig.\
\ref{fig3} we compare the measured shift with a model in which we
set the collision energy of the particles to be the Fermi energy
plus the ground state energy. We numerically calculate the Fermi
energy for the noninteracting gas in the full three-dimensional
configuration of the optical lattice and the harmonic confining
potential. We use a tight-binding model for the direction of the
lattice laser and a harmonic oscillator potential in the
transverse directions. Using the parameterization of the Feshbach
resonance according to \cite{Ticknor2004}, we obtain the shifted
position of the resonance for a given lattice depth. For the
$|m|=1$ branch of the resonance we find good agreement of our data
with the theory whereas for the $m=0$ branch the observed shift is
larger than predicted by our model. There may be an additional
confinement induced shift of the $p$-wave resonance which depends
on the $m$-quantum numbers in the collision process
\cite{Granger2004}, however no quantitative theory is available.
The observed increasing width of the Feshbach resonance is also
due to a larger collision energy \cite{Ticknor2004}.

% 1D Fermi system - principle
Reducing the dimensionality further, we study the effect of the
alignment of the electronic spins on the $p$-wave interaction in a
one-dimensional quantum gas. Therefore all spins are lined up
either orthogonal (see Fig.\ \ref{fig1}b) or parallel (Fig.\
\ref{fig1}c) to the orientation of the gas. We prepare the
one-dimensional Fermi gases by superimposing a second standing
wave laser field onto the two-dimensional quantum gases
\cite{Moritz2005}. Either the $x$- or the $y$-direction of the
optical dipole trap is slowly turned off and replaced by an
optical lattice along the same direction and
having the same beam geometry.% \cite{rampslattice}

\begin{figure}[htbp]
  \includegraphics[width=.77\columnwidth,clip=true]{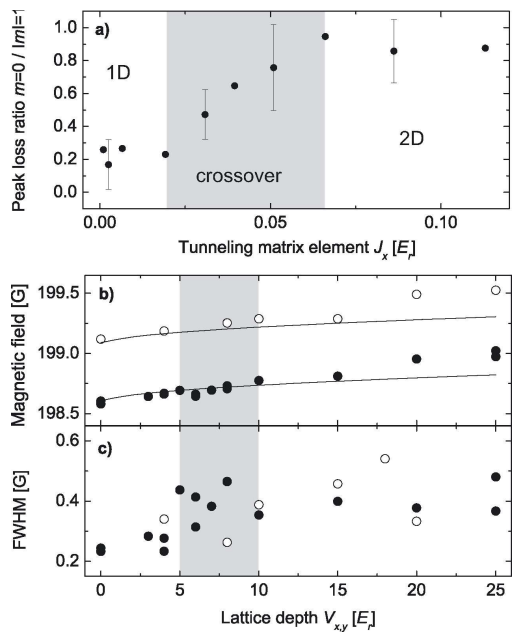}
  \caption{a) Suppression of collisional losses in the $m=0$ partial wave
  versus the tunneling matrix element between the one-dimensional quantum
  gases. The bars reflect the errors determined from the fit to the loss
  peaks of two measurements. The grey box indicates
  the crossover region from a 2D to a 1D quantum gas as inferred from the inhibition of transverse collisions.
  b) Shift of the position of the Feshbach resonance  when
  tuning the gas from 2D to 1D. Open symbols indicate the position of the $m=0$
  branch in the $y$-$z$-lattice (see Fig.\ \ref{fig1}c) and the solid symbols the $|m|=1$ branch of the resonance in the
  $x$-$z$-lattice (see Fig.\ \ref{fig1}b).
  The solid lines show a calculation of the expected positions (see text). c) Width of the Feshbach
  resonance loss feature.}
  \label{fig4}
\end{figure}

% 1D Fermi gas - suppression of losses
We now consider the orthogonal configuration where only collisions
with $|m|=1$ are possible, and correspondingly we observe only
this branch of the Feshbach resonance (see Fig.\ \ref{fig2}c). To
study the suppression of the $m=0$ branch quantitatively we create
a two-dimensional optical lattice along the $x$- and the
$z$-direction with $V_z=25\,E_r$ and adjustable $V_x$. We have
measured the peak loss on the $m=0$ and the $|m|=1$ resonance
position, respectively. In Fig.\ \ref{fig4}a we plot the ratio of
the peak loss versus the tunneling matrix element along the
$x$-direction, i.\,e.\ between the tubes of the optical lattice.
For no tunneling the one-dimensional gases are well isolated and
losses on the $m=0$ branch are completely suppressed. For larger
tunneling rates hopping of atoms between the tubes is possible and
the system is not kinematically one-dimensional anymore but in a
crossover regime. Therefore collisions in the $m=0$ branch become
possible which give rise to losses. The measurement directly
verifies suppressed tunneling between neighboring lattice tubes
and proves that the gases in the individual lattice tubes are
kinematically one-dimensional. Orienting the one-dimensional
quantum gases parallel to the magnetic field axis, we observe the
$m=0$ branch of the Feshbach resonance only (see Fig.\
\ref{fig2}d). Note that the width and the position of crossover
regime is determined by the time scale of the physical processes
under investigation: in this paper we are one-dimensional with
respect to atomic collision time scales but not necessarily for
slower dynamical processes such as collective oscillations
\cite{Moritz2003}.

% 1D shift.
For the one-dimensional Fermi gases we observe a further shift of
the resonance position and a broadening of the loss feature as
compared to the higher-dimensional configurations. A confinement
induced shift of the $p$-wave resonance in one dimension for the
$m=0$ branch has been predicted in addition to the increased
ground state and Fermi energy \cite{Granger2004}. We apply the
calculation technique as described above to obtain the Fermi
energy of the gas and use the same parameterization of the
Feshbach resonance. For the $m=0$ branch we additionally include
the theory of ref.\ \cite{Granger2004} for all lattice depths.
This introduces a small error in the 2D and crossover regime. We
note, however, that the confinement induced shift is small as
compared to the shift due to the increased collision energy. A
comparison between the resulting shift and the experimental data
is shown in Fig.\ \ref{fig4}b. The increasing width of the loss
feature is expected because the width of the Feshbach resonance
also depends on the energy of the particles involved in the
collision process \cite{Ticknor2004}.

% 3D lattice
By using three orthogonal standing waves, we prepare a
band-insulating state in a 3D optical lattice \cite{Koehl2005}
where the atoms are localized in the potential wells with at most
one atom per lattice site. In this ''zero-dimensional`` situation
all $p$-wave scattering is completely inhibited (see Fig.\
\ref{fig2}e) and no loss features are observed.

% Conclusions
In conclusion, we have studied spin-polarized interacting Fermi
gases in low dimensions using a $p$-wave Feshbach resonance. We
demonstrate that in reduced dimensions the direction of
spin-alignment significantly influences the scattering properties
of the particles. Moreover, we find a confinement induced shift of
the resonance position and observe good agreement with a
theoretical model. Strongly interacting low-dimensional Fermi
gases offer a wealth of fascinating many-body phenomena
\cite{Ho2005}.

%The prospect of $p$-wave superfluidity \cite{Anderson1961} appears
%very intriguing in cold atomic gases \cite{Ho2005,Cheng2005} and
%especially in two-dimensional systems exotic phases are predicted
%\cite{Gurarie2005,Iskin2005} where the spin-alignment is
%essential.

We would like to thank D.\ Blume for discussions, and SNF and SEP
Information Sciences for funding.


\begin{thebibliography}{}
\bibitem{shift} We have observed the opposite systematic shift when reversing the direction of the final sweep.
\bibitem{calibrationlattice} The lattice depth was calibrated by intensity modulation and studying the parametric heating.
The calibration error is estimated to be $<10\%$.
\end{thebibliography}
\end{document}